\documentstyle[prd,aps,preprint,epsf,epsfig]{revtex}
\firstfigfalse
\begin{document}
\title{ 
 Gauged Nambu-Jona Lasinio Studies of the Triviality of Quantum Electrodynamics}
\vskip -1 truecm

\author{S. Kim}
\address{Department of Physics, Sejong University,
Seoul 143-747, Korea}

\author{John~B.~Kogut }
\address {Physics Department, University of Illinois at Urbana-Champaign,
Urbana, IL 61801-30}

\author{ Maria--Paola Lombardo}
\address{Istituto Nazionale di Fisica Nucleare,
         Sezione di Padova, e Gr. Coll. di Trento, Italy}

\date{\today}
\maketitle
\vskip -1 truecm

\begin{abstract}
By adding a small, irrelevant four fermi interaction to the action of noncompact lattice Quantum
Electrodynamics (QED), the theory can be simulated with massless quarks in 
a vacuum free of lattice monopoles. Simulations directly in the chiral limit of massless
quarks are done with high statistics on $12^4$, $16^4$ and $20^4$ lattices
at a wide range of couplings with good control over finite size effects, systematic and
statistical errors. The lattice theory possesses a second order 
chiral phase transition which we show is logarithmically trivial, with the same
systematics as the Nambu-Jona Lasinio model. The irrelevance of the four
fermi coupling is established numerically. Our fits have excellent 
numerical confidence levels. 
The widths of the scaling windows are examined in both
the coupling constant and bare fermion mass directions in parameter space. For vanishing
fermion mass we find a broad scaling window in coupling which is essential
to the quality of our fits and conclusions. By adding a small
bare fermion mass to the action we find that the width of the scaling window 
in the fermion mass direction is very narrow. Only when a subdominant scaling term 
is added to the leading term of
the equation of state are adequate fits to the data possible. The failure of past studies
of lattice QED to produce equation of state fits with adequate confidence levels
to seriously address the question of triviality is explained. 
The vacuum state of the lattice model is probed for topological excitations, such
as lattice Monopoles and Dirac strings, and these objects are shown to be non-critical along
the chiral transition line as long as the four fermi coupling is nonzero. 
Our results support Landau's contention that perturbative QED suffers from 
complete screening and would have a vanishing fine structure constant 
in the absence of a cutoff.
 
\end{abstract}

\pacs { 
12.38.Mh,
12.38.Gc,
11.15.Ha
}
\newpage

\section{Introduction}


Simulation studies of Nambu-Jona Lasinio models have proven to be much more quantitative
than those of other field theories \cite{Looking}. In particular, the logarithmic
triviality of these models has been demonstrated, although determining logarithmic
singularities decorating mean field scaling laws is a daunting
numerical challenge. The reason for this success lies in the fact that when one formulates
these four fermi models in a fashion suitable for simulations, one introduces an auxiliary scalar
field $\sigma$ in order to write the fermion terms of the action as a quadratic form. In this
formulation $\sigma$ then acts as a chiral order parameter which receives a vacuum
expectation value, proportional to the chiral condensate $<\bar\psi \psi>$, in
the chirally broken phase. Most importantly,
the auxiliary scalar field $\sigma$ becomes the dynamical mass term in the
quark propagator. The Dirac operator is now not singular for quarks with vanishing
bare mass and its inversion \cite{HMC}, \cite{HMD} is
successful and very fast. The algorithm for Nambu-Jona Lasinio models is
"smart" --- it incorporates a
potential feature of the solution of
the field theory, chiral symmetry breaking and a dynamical fermion mass, 
into the field configuration generator.

The good features of the simulation algorithm for the  Nambu-Jona Lasinio model
can be generalized to lattice QCD \cite{KD} and QED \cite{PLB} by incorporating
a weak four fermi term in their actions. These generalized models now depend on two couplings,
the familiar gauge coupling and a new four fermi coupling. By choosing the four fermi
coupling small we can be confident that all the dynamics resides in the gauge
and fermi fields and the four fermi term just provides the framework for an improved algorithm
which allows us to simulate the chiral limit of massless quarks directly.

We shall find a line of spontaneously broken chiral symmetry transition points
in the two dimensional coupling constant parameter space of the U(1)-gauged
Nambu-Jona Lasinio model. By
simulating the model at several regions along the transition line, we will
see that the theory is logarithmically trivial and that the four fermi term is irrelevant
in the continuum limit. Our conclusions will be supported by fits with very high confidence
levels. Because of the irrelevance of the pure
four fermi interaction, this model will make "textbook" QED
accessible and this paper will address the classic problem
of whether QED suffers from complete charge screening. Our
measurements will show that the theory is logarithmically trivial and
the systematics of the logarithms of triviality follow those of
the Nambu-Jona Lasinio model rather than the scalar $\phi^4$ model as usually assumed.

Simulating the $m=0$ case directly has substantial advantages, both theoretical and practical.
When $m$ is set to zero, the theory has the exact chiral symmetry of the interaction
terms in the action and this forbids chiral symmetry breaking counterterms from appearing
in its effective action. This simplicity can lead to a large scaling window in
the direction of the gauge or four fermi coupling in the theory's
parameter space. Our simulation results will support this point. However, when $m$ is
not zero, as in most past studies of lattice QED and QCD, the effective action has
no protection from dangerous symmetry breaking counterterms. In fact we will find
that the scaling window of the lattice theory in the $m$-direction is very small and
this fact is responsible for the failure of past approaches to lattice QED to 
address the question of triviality in a straightforward, convincing fashion.
In fact, \cite{Kogut,HKKperc} claimed non-triviality for the theory while
\cite{Goc,referee4} found triviality and backed up their claim further in
\cite{Goc} by calculating the sign of the beta function, which is directly relevant
to the question of triviality.

In addition, we shall check that the algorithm used in this work generates 
gauge field configurations for couplings near the chiral transition line which
are free of lattice artifacts, such as monopoles \cite{HW} and Dirac strings, etc.

In this paper we will present data and analyses. Preliminary results have
already appeared in letter form \cite{PLB}, but this article will contain new data,
analyses and discussions. Other applications of the use of a four fermi
term to speed lattice gauge theory simulations are also under development and
are being applied to QCD \cite{KD}. It is important to note that in these applications
the strength of the four fermi term is weak, so it is not responsible for chiral 
symmetry breaking. It just acts as scaffolding which leads to an algorithm
that converges efficiently in the limit of massless quarks. The dynamics
resides in the gauge and fermion field interactions.

This paper is organized as follows. In the next section we present the formulation
of the lattice action and discuss its symmetries and general features. In the third section
we test the algorithm and tune its parameters. In the next three sections we present data
and analyses over a range of gauge couplings for three choices of the irrelevant 
four fermi coupling on $16^4$ lattices. The irrelevance of the four fermi coupling is
demonstated explicitly and equation of state fits are presented which show that the 
theory is logarithmically trivial with the same systematics as the Nambu-Jona Lasinio model.
The confidence levels of these fits range from approximately $35$ to $98$ percent.
Analyses of the order parameter's susceptibility reinforce our conclusions. In the seventh
section we consider simulations at nonzero bare fermion masses in order to make contact
with past work on pure lattice QED. We find that subdominant scaling terms are needed to fit
the data. In other words, the usual assumption that the scaling window is wide enough to address
the issue of triviality by simulating the model at nonzero fermion masses and fitting to
a logarithmically improved mean field form is shown to be incorrect. In section eight we present data
on lattices ranging in size from $12^4$ to $20^4$ to check that our data for the 
chiral condensate is not influenced significantly by finite size effects for the range of
couplings used in the fits. 
In section nine we consider measurements of lattice
monopole observables to check that they are not critical at the chiral transition points as
long as the bare four fermi coupling is nonzero. In section ten we discuss the possible role of
lattice artifacts in simulations of pure lattice QED and address some concerns in the literature.
Finally, in section eleven we suggest additional work in this field.

\section{Formulation}
We considered the $U(1)-$gauged Nambu Jona Lasinio model with four
species of fermions. The Lagrangian for the continuum Nambu-Jona Lasinio model is,

\begin{equation}
L  = \bar \psi (i\gamma \partial -e \gamma A - m) \psi -
\frac{1}{2}G^2 (\bar \psi \psi) ^2 -\frac {1}{4} F^2
\end{equation}

The symmetries and other properties of $L$ have been discussed in \cite{PLB} and
we refer the reader to that and related references for details. We will be
brief here and just review a few conceptually important points.

The pure Nambu-Jona Lasinio model (Eq. 1. with $e$ set to zero) has been solved at large $N$ by
gap equation methods \cite{RWP}, and an accurate simulation study of it has been
presented \cite{Looking}.

The lattice Action for Eq. 1 reads:

\begin{equation}
S  =  \sum_{x,y} \bar\psi(x) (M_{xy} + D_{xy}) \psi(y) + 
  \frac {1}{2 G^2} \sum_{\tilde x} \sigma ^2 (\tilde x) +
  \frac{1}{2 e^2} \sum_{x,\mu,\nu} F^2_{\mu\nu}(x)
\end{equation}

\noindent
where 
\begin{eqnarray}
F_{\mu\nu}(x)& = &\theta_{\mu}(x) + \theta_{\nu}(x+\hat{\mu}) +
\theta_{-\mu}(x+\hat{\mu}+\hat{\nu}) + \theta_{-\nu}(x+\hat{\nu}) \\
M_{xy}& = & (m + \frac{1}{16} \sum_{<x,\tilde x>} \sigma( \tilde x))
\delta_{xy} \\
D_{xy} & = & \frac{1}{2} \sum_\mu \eta_{\mu} (x) (
 e^{i\theta_{\mu}(x)} \delta_{x+\hat{\mu}, y}
- e^{-i\theta_{\mu}(y)} \delta_{x-\hat{\mu}, y} )
\end{eqnarray}

\noindent
where $\sigma$ is an auxiliary scalar field defined on the sites of the dual
lattice $\tilde x$ \cite{CER}, and the symbol $<x,\tilde x>$ denotes
the set of the 16 lattice sites surrounding the direct site $x$.
The factors $e^{\pm i\theta_\mu}$ are the gauge connections and
$\eta_\mu(x)$ are the staggered phases, the lattice analogs of the
Dirac matrices. $\psi$ is a staggered fermion 
field and  $m$ is the bare fermion mass, which will be set to 0.
Note that the lattice expression for $F_{\mu\nu}$ is non-compact
in the lattice field $\theta_{\mu}$, while the gauge field couples to the
fermion field through compact phase factors which guarantee local gauge
invariance. This point will be discussed further in Sec.10 below.

Interesting limiting cases of the above Action are the pure $Z_2$ Nambu-Jona Lasinio
model ($e=0$), which has a phase transition at $G^2 \simeq 2$ \cite{Looking} and
the pure lattice QED ($G=0$) limit, whose chiral phase transition is
near $\beta_e \equiv 1/e^2 = .204$ for four flavors \cite{Kogut}, \cite{Az}.
The pure QED ($G=0$) model also has a monopole percolation transition
which is probably coincident with its chiral transition at $\beta_e = .204$ \cite{KW}.
Past simulations of this lattice model have led to contradictory results    
\cite{Az}, \cite{Goc}. 
Since the gauged Nambu-Jona Lasinio model can be simulated at $m=0$ for all gauge
couplings, the results reported here will 
be much more precise and decisive than those of the pure lattice QED ($G=0$) limit. 

\begin{figure}

\epsfig{file=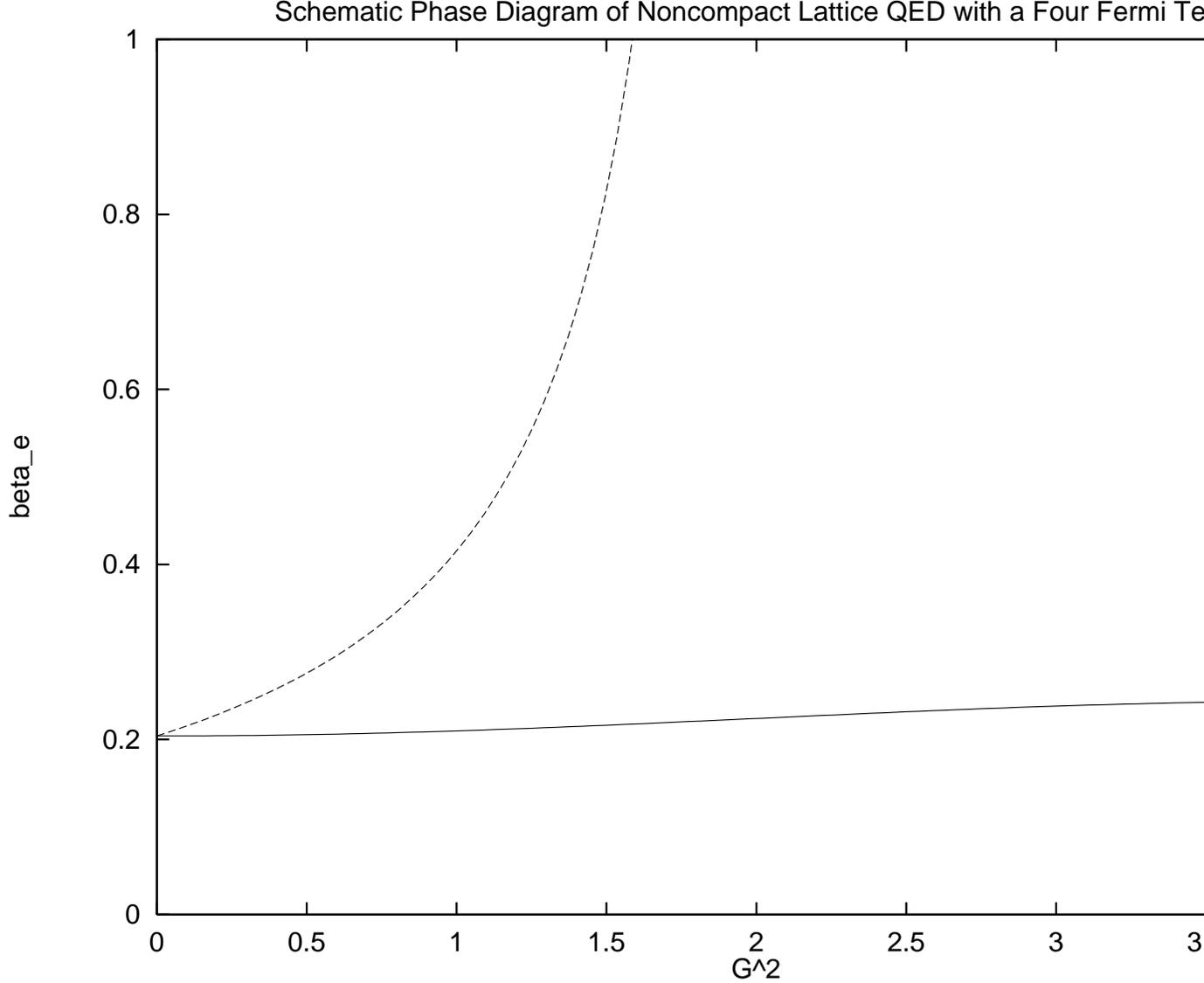,width=15cm,angle=-90,clip=}
\vspace{0.3cm}
\caption{The upper, dashed line labels chiral transitions and the lower, 
solid line labels monopole percolation transitions.}
\label{fig:z2gph}
\end{figure}

We scanned the 2 dimensional parameter space ($\beta_e$,$G^2$) using the Hybrid Molecular Dynamics 
algorithm tuned for four continuum fermion species \cite{HMD}
and measured the chiral condensate and monopole susceptibility as
a function of $\beta_e$ and $G^2$.
We found that as we increased $G^2$ and 
moved off the $G=0$ axis, the peak of the monopole susceptibility shifted 
from $\beta_e = .204$ at $G=0$ to $\beta_e = .244$ at large $G$.
By contrast the chiral transition point shifted to a larger $\beta_e$ 
than the monopole percolation transition for a given value of $G$
and became distinct from the monopole percolation point as soon as $G$ became 
nonzero, as shown in the Phase Diagram, Fig.1.

\section{Controlling Systematic $dt$ Errors in the Algorithm.}

Before turning to physically interesting measurements, we should address some technical issues
concerning the algorithm. Unlike the Hybrid Monte Carlo algorithm, the Hybrid
Molecular Dynamics algorithm is not exact. The molecular dynamics equations of motion
can be found in the literature \cite{Dag}. In order to evolve
the noisy equations of motion and generate an ensemble of field configurations,
one must choose a Monte Carlo time step $dt$ \cite{HMD}. 
The discretization errors have been exhaustively studied and it has been
shown that systematic errors in observables behave as $dt^2$ \cite{Duane}. Therefore, we must choose $dt$
small enough that these systematic errors are no larger than the statistical errors
we will encounter.

In Table 1 and Fig. 2 we show the order parameter $\sigma$ evaluated on a $12^4$ lattice
at gauge coupling $\beta_e = 0.25$ and four fermi coupling $G^2 = 1/4$. (We write $\sigma$ here 
as a shorthand for $<\sigma>$, the expectation value of the field. This is a standard notational shortcut
which, hopefully, shouldn't lead to confusion.) The table shows that
as long as $dt < 0.03$ the systematic error in $\sigma$ is negligible. The figure shows that
the theoretically expected quadratic dependence of the systematic error on $dt^2$ 
has been confirmed numerically.
The error bars quoted in the table have been obtained using the usual binning techniques, so they
reflect the correlations in the measurements. The last column of the table gives the number
of trajectories in each data set. A trajectory here means an interval of one Monte Carlo time
unit of the algorithm ( for $dt=0.01$ a trajectory consists of one hundred sweeps ). After each
trajectory a single measurement of $\sigma$ was made.

Most of our production runs were done using $dt = 0.02$. Particularly close to the critical
point where these systematiic errors are most dangerous, we checked our results with runs
having $dt = 0.01$. No problems were encountered.

\begin{figure}
\epsfig{file=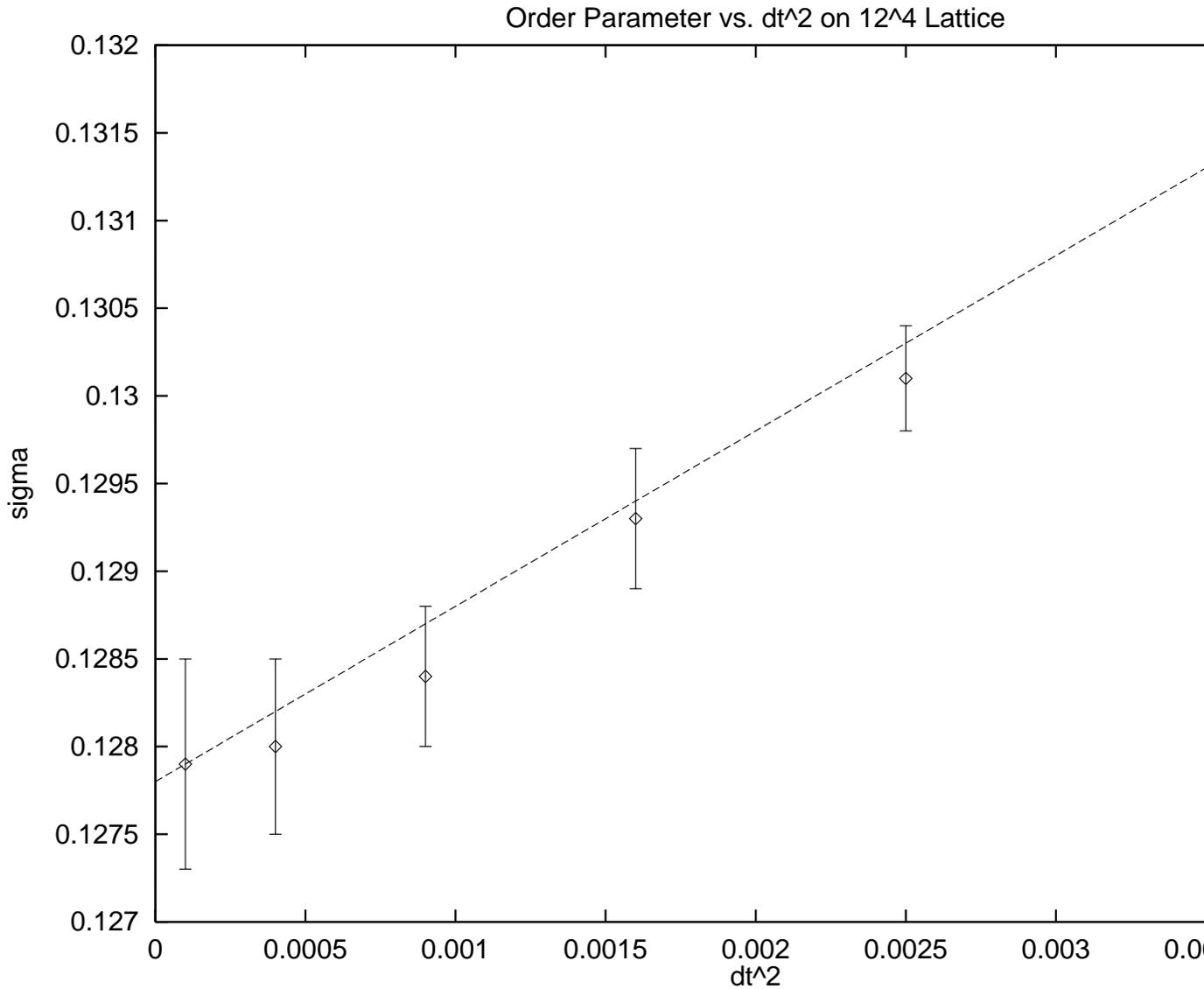,width=15cm,angle=-90,clip=}
\vspace{0.3cm}
\caption{$\sigma$ vs. $dt^2$}
\label{fig:z2gdt2}
\end{figure}

\section{Simulations at $G^2 = 1/4$ on a $16^4$ lattice.}

As stated in the Introduction, we made accurate measurements on the chiral critical line 
for many choices of
couplings ($\beta_e$, $G^2$) and lattice sizes ranging from $12^4$ to $20^4$. In this section
we review our data collected varying $\beta_e = 0.15-.30$
at fixed $G^2 = 1/4$ on a $16^4$ lattice. A discussion and presentation of this data
has appeared in \cite{PLB}, so we will be brief.

The data is presented in Table 2. The columns list the average values of $\sigma$, $\chi_{\sigma}$
which is the longitudinal susceptibility of the order parameter \cite{Koc}, $M$ which is 
the monopole percolation order parameter and $\chi_M$ which is the 
susceptibility of the monopole order parameter \cite{HW}. The monopole
observables will be discussed later.

The data for the order parameter was fit to a form which could accomodate either $\phi^4$
or Nambu-Jona Lasinio triviality:
$\beta_c-\beta_e = a \sigma^{2}\ln^p(b/\sigma)$, where the parameter $p$, the critical point
$\beta_c$, the amplitude $a$ and the scale $b$ are determined by the 
fitting routine. Recall that $\phi^4$ triviality gives $p=-1$ and Nambu-Jona Lasinio triviality
gives $p=+1$. For the scaling window of
gauge couplings $\beta_e$ between $.18$ and $.225$, we found 
the parameters $\beta_c = .2350(1)$, $a = 34.3(3.9)$, $\ln b = 1.55(10)$ and
$p = 1.00(8)$ with a confidence level of 34 percent. The reader should consult
the figures and discussison in \cite{PLB} for more detail and perspective.
As will be discussed below in Sec.9, these simulations also
measured topological observables for the system's
vacuum and we confirmed that monopoles and related objects were \underline{not} critical near
the chiral transition $\beta_c = .2350(1)$, $G^2 = 1/4$. ( We shall see there that 
the monopole percolation transition is very narrow and occurs at $\beta_e = .2175(25)$ for $G^2 = 1/4$. )

Are other fitting forms possible for this data? This is certainly true, of course. The point we are making,
however, is that log-improved mean field theory fits the data with very high confidence levels and
there are compelling theoretical reasons for it. The data and
the fits support the 'conventional wisdom' that QED is a trivial field theory and that
the logarithms of triviality follow the systematics of the Nambu-Jona Lasinio model rather than
the scalar $\phi^4$ model. This last point is different from that usually assumed. 
In retrospect, it is very plausible that the
Nambu-Jona Lasinio model represents the triviality of QED better than $\phi^4$, but the differences
between the two models have not been emphasized or appreciated in the past.

Let's end our discussion of $\sigma$ with some examples of other fits. Simple power laws are the first
ones to try. For example, a fit of the form $\sigma = a (\beta_c-\beta_e)^{\beta_{mag}}$ is expected
to work rather well with $\beta_{mag}$ slightly larger than $1/2$ since the Nambu-Jona Lasinio fits
have worked so well. If we choose the range of $\beta_e$ to extend from $.15$ to $.225$, we find 
$\beta_{mag} = 0.530(6)$, $a = 0.449(3)$, $\beta_c = 0.23315$, but the confidence level is very
poor, ($\chi^2/d.o.f. \approx 113/8$). If we accept only a smaller range of couplings closer to the
critical point, $\beta_e$ extending from $.18$ to $.225$, the quality of the fit improves
($\chi^2/d.o.f. \approx 6.8/5$, confidence level of $24$ percent) while the 
critical index $\beta_{mag}$ rises to $0.576(12)$. This
is the trend we find in the data : powerlaw estimates of the critical index $\beta_{mag}$ increase
as the range of couplings is restricted closer and closer to the critical coupling. This systematic
drift in the fitting results suggests that a simple power is not an adequate representation of the full
data set, but is simply mocking up the logarithm of the Nambu-Jona Lasinio fit, which has a higher confidence
level and is stable as different ranges of $\beta_e$ are considered.

In \cite{PLB} we also analyzed the susceptibility associated with $\sigma$.
In mean field theory,
the singular piece of the longitudinal susceptibility $\chi$ 
diverges at the critical point $\beta_c$ as $\chi_+ = c_+ |t|^{-\gamma}$, 
$t \equiv (\beta_c-\beta_e)/\beta_c$, as $t$ approaches zero from above in the 
broken phase, and as $\chi_- = c_- |t|^{-\gamma}$ in the 
symmetric phase \cite{ID}. The critical index $\gamma$ is exactly unity in mean field theory.

In logarithmically trivial models $\gamma$ remains unity, but the amplitudes
$c_+$ and $c_-$ develop weak logarithmic dependences \cite{ID}. In the 
two component $\phi^4$ model, $c_-/c_+ = 2 + \frac{2}{3}/\ln(\frac{b}{\sigma})$, while 
in the $Z_2$ Nambu-Jona Lasinio model, $c_-/c_+ = 2 - 1/\ln(\frac{b}{\sigma})$ 
\cite{Looking}, where the scale $b$ comes from the order parameter fit. 
Constrained linear fits to
the data \cite{PLB} produced the amplitude ratio $c_-/c_+ = 1.74(10)$.
Since $\sigma$ varies from $.0953(1)$ to $.0367(2)$ over the $\beta_e$ range $.18$ - $.225$ of the
scaling window in the broken phase, the logarithm in the theoretical prediction 
of the Nambu-Jona Lasinio model states that
$c_-/c_+$ should range from $1.75$ to $1.79$. Again, the 
agreement between the simulation data and theory is very good.

We find no support for the approximate analytic schemes discussed in \cite{Az3}
which predicted that gauged $U(1)$ Nambu-Jona Lasinio models with a four fermi term with continuous
chiral symmetry are nontrivial, have powerlaw critical singularities with indices that vary
continuously with the couplings $\beta_e$ and $G^2$. Additional
simulations in Sec.5 and 6 below will give strong evidence for the irrelevance of the four fermi term
contrary to the results of \cite{Az3} .
The reader should recall that truncated $U(1)$ Nambu-Jona Lasinio models which account for 
only restricted sets of Feynman diagrams produce nontrivial "theories" with
critical indices that vary continuously as $\beta_e$ and $G^2$ are varied. 
For example, this occurs if only "rainbow graphs" of gauged 
Nambu-Jona Lasinio models are summed \cite{Love}. Some of these exercises may be relevant
to Technicolor model building.

On the basis of the work here, however, we suspect that when fermion vacuum polarization is
accounted for, one would find complete charge screening and every gauged Nambu-Jona Lasinio model
based on continuum noncompact $U(1)$ gauge dynamics would be trivial for all couplings. We suspect that
nontriviality and lines of nontrivial field theories are aspects of truncation
procedures only. We suspect, on the basis of the present work and past
triviality investigations in scalar QED \cite{Baig}, that only models with dynamics 
beyond continuum noncompact $U(1)$
gauge fields and fermions can be nontrivial and have a renormalization group 
fixed point at nonzero gauge coupling. An example might be afforded by $U(1)$
theories with fundamental monopoles \cite{HK}.

\section{Simulations at $G^2 = 1/8$ on a $16^4$ lattice.}

In this section
we consider new data collected varying $\beta_e = 0.16-.25$
at fixed $G^2 = 1/8$ on a $16^4$ lattice. 

The purpose of this series of simulations is to 1. to verify that the four fermi coupling is irrelevant, and 2. to
accumulate more evidence that the theory is logarithmically trivial in the sense of the
Nambu-Jona Lasinio model.

The data is presented in Table 3 in the same format as Table 2.

In Fig. 3 we show the data for the chiral condensate $\sigma$,
at fixed $G^2 = 1/8$ and variable $\beta_e$. We use the same fitting procedures as
used in Sec.4 :
$\beta_c-\beta_e = a \sigma^{2}\ln^p(b/\sigma)$, where the parameter $p$, the critical point
$\beta_c$, the amplitude $a$ and the scale $b$ are determined by the 
fitting routine. For the scaling window of
gauge couplings $\beta_e$ between $.17$ and $.205$, we found 
the parameters $\beta_c = .21470(5)$, $a = 12.02(1.18)$, $\ln b = 0.40(10)$ and
$p = 1.07(8)$ with a confidence level of 87 percent. This excellent fit is
the one shown in the figure. Note that eight data points for $\beta_e$ between $0.17$ and $0.205$
were used in the fit while the figure has two additional points at stronger coupling. Those points lie slightly
below the fit, are slightly outside the scaling window and show the extent of the scaling window.


\begin{figure}
\epsfig{file=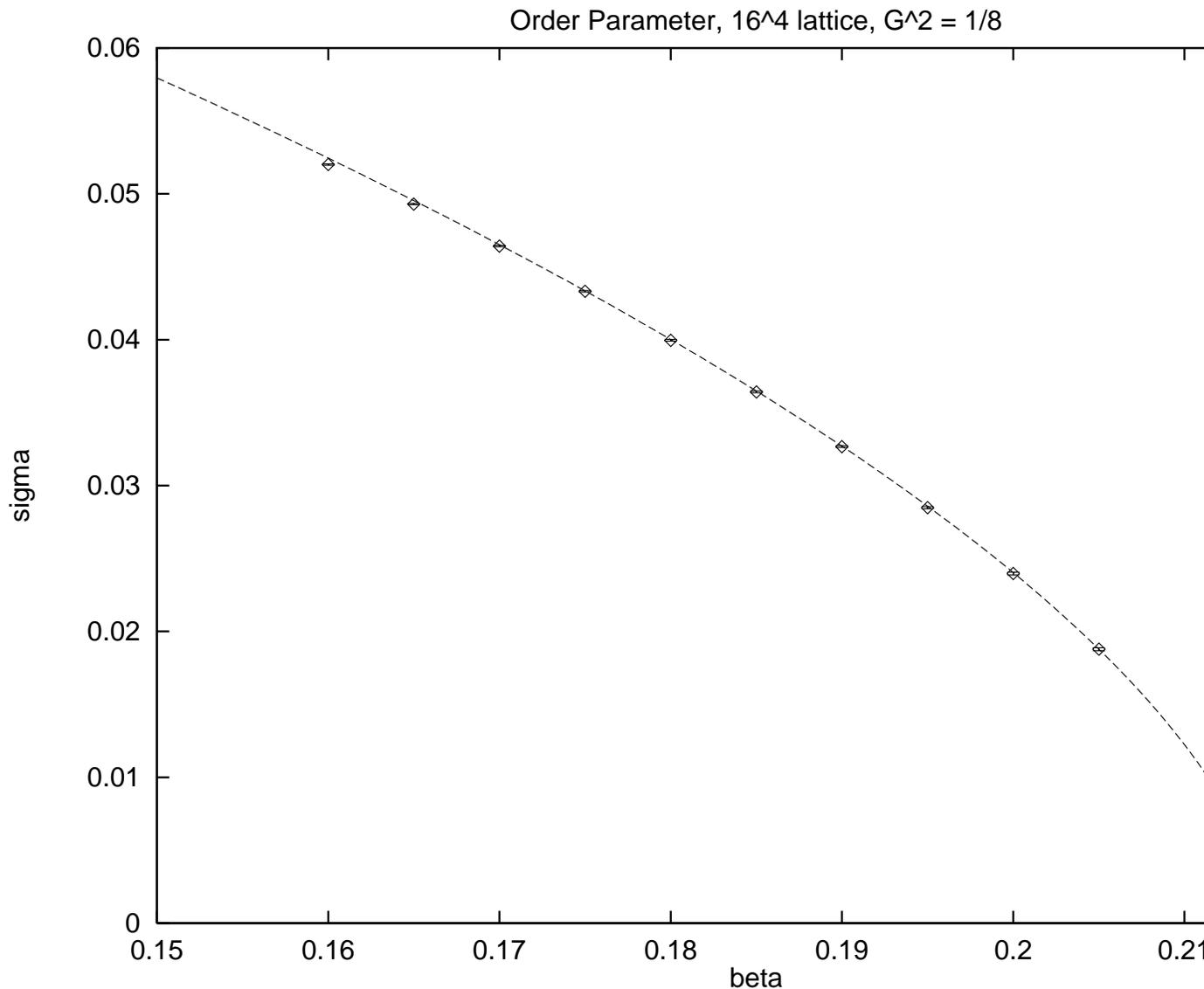,width=15cm,angle=-90,clip=}
\vspace{0.3cm}
\caption{$\sigma$ vs. $\beta_e$ for $G^2 = 1/8$}
\label{fig:mag1616}
\end{figure}

We plot the eight data points between $\beta_e$ of $0.17$ and $0.205$ and the fit as shown in
Fig.4., $|\beta_c-\beta_e|/\sigma^2$ vs. $\ln(1/\sigma)$ to illustrate the importance
and numerical significance of the logarithm. 
The dashed line is the previous fit redrawn in this format, where we have 'zoomed' in on the scaling
window for emphasis. Clearly this fit is stable to further cuts on the data set
since all the data points lie on the fit.


\begin{figure}
\epsfig{file=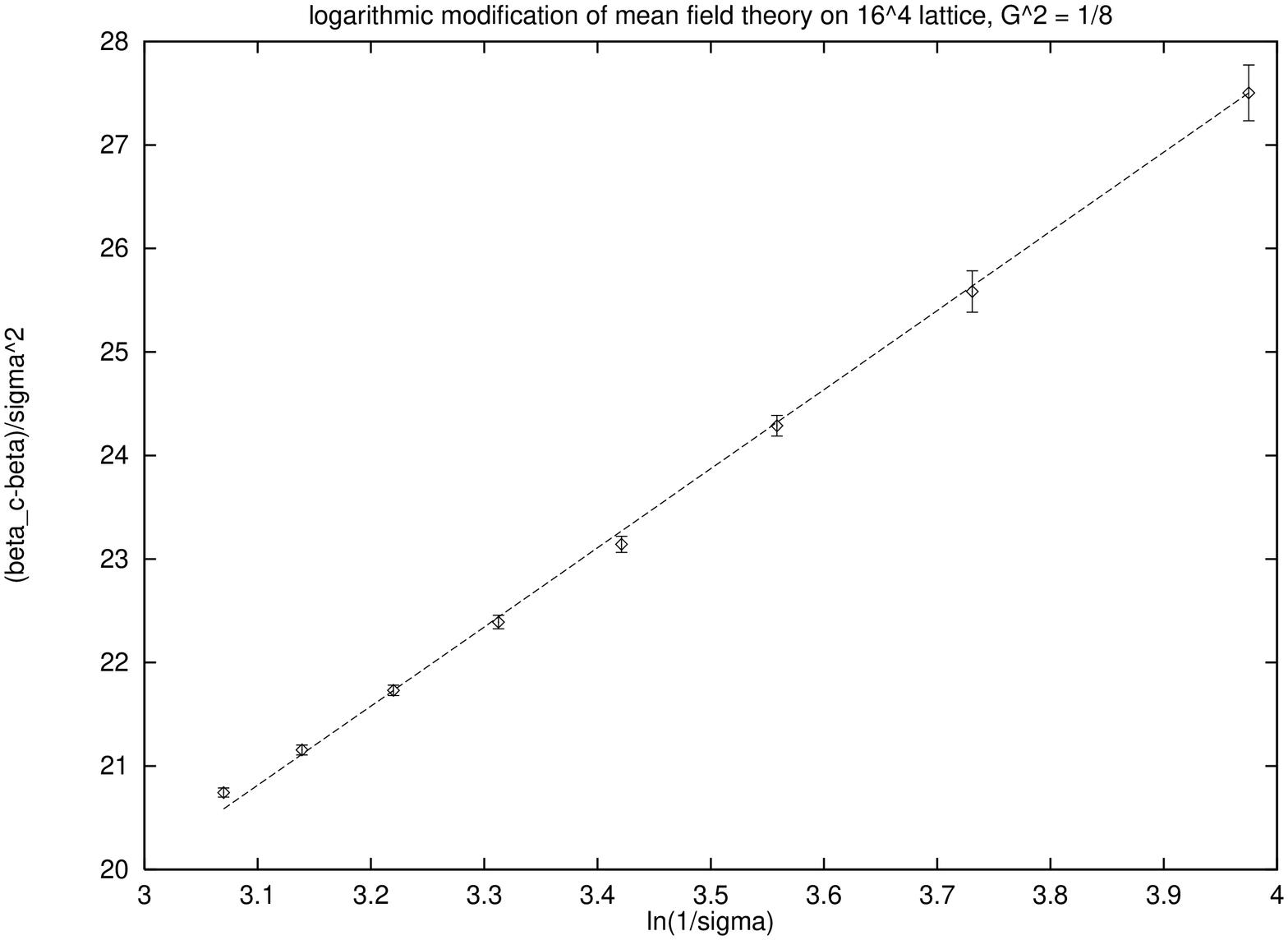,width=15cm,angle=-90,clip=}
\vspace{0.3cm}
\caption{$|\beta_c-\beta_e|/\sigma^2$ vs. $\ln(1/\sigma)$ for $G^2 = 1/8$}
\label{fig:sigz2ga}
\end{figure}

We conclude that Nambu-Jona Lasinio triviality accommodates the lattice data at $G^2 = 1/8$
with very good confidence levels.
This success also shows the irrelevance of the four fermi term in the lattice action : the
scaling law for the order parameter is the same as that at the larger $G^2$ value although 
the lattice parameters, such as the location of the critical point, have changed.

\begin{figure}
\epsfig{file=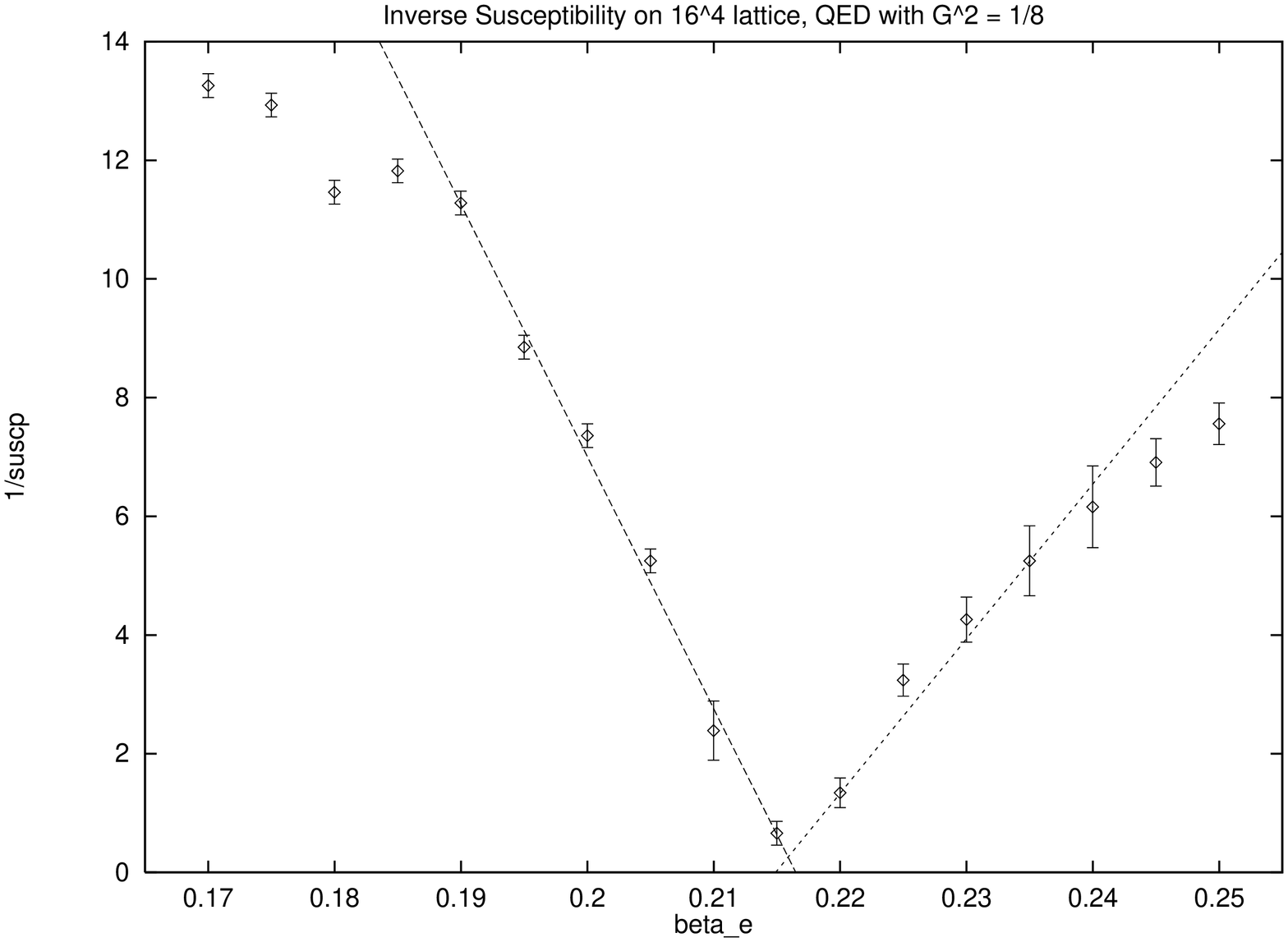,width=15cm,angle=-90,clip=}
\vspace{0.3cm}
\caption{Inverse Susceptibility vs. Coupling $\beta_e$, $G^2 = 1/8$}
\label{fig:insusz2g1616}
\end{figure}

Next, in Fig.5 we show the inverse of the
longitudinal susceptibility of the auxiliary field $\sigma$ 
at fixed $G^2 = 1/8$ and variable $\beta_e$. We follow the same procedures as used in Sec.4
to analyze and plot the data here.
The plot picks out a critical point $\beta_c = .2155(10)$ and is consistent with 
the mean field value of the critical index $\gamma = 1.0$.
The constrained linear fits determine the amplitude ratio, $c_-/c_+ = 1.65(10)$. 
Since $\sigma$ varies from $.04642(5)$ to $.01878(10)$ over the $\beta_e$ range $.17$ - $.205$ of the
scaling window in the broken phase, the logarithms in the theoretical prediction 
of the Nambu-Jona Lasinio model for the amplitude ratio predict that
$c_-/c_+$ should range from $1.75$ to $1.79$. Again, the 
agreement between the simulation data and theory is good, but is not comparable in
quality or decisiveness to our other fits.

\section{Simulations at $G^2 = 1/2$ on a $16^4$ lattice.}

In this section
we consider new data collected varying $\beta_e = 0.17-.37$
at fixed $G^2 = 1/2$ on a $16^4$ lattice. In this case the four fermi coupling
is four times stronger than the data discussed in the previous section, but far too weak to
cause chiral symmetry breaking in the absence of the gauge coupling.

The analysis and plots here are identical to the previous discussions of $G^2 = 1/4$ and 
$G^2 = 1/8$, so we will be brief.

The data is presented in Table 4 in the same format as Table 1.

In Fig.6 we show the data for the chiral condensate $\sigma$,
at fixed $G^2 = 1/2$ and variable $\beta_e$. We use the same fitting procedures as
used in Sec.3 :
$\beta_c-\beta_e = a \sigma^{2}\ln^p(b/\sigma)$, where the parameter $p$, the critical point
$\beta_c$, the amplitude $a$ and the scale $b$ are determined by the 
fitting routine. For the scaling window of
gauge couplings $\beta_e$ between $.22$ and $.27$, we found 
the parameters $\beta_c = .29117(5)$, $a = 20.0(4.1)$, $\ln b = 1.6(4)$ and
$p = 0.86(18)$ with a confidence level of 99.9 percent. This impressive fit is the one
shown in the figure.

\begin{figure}
\epsfig{file=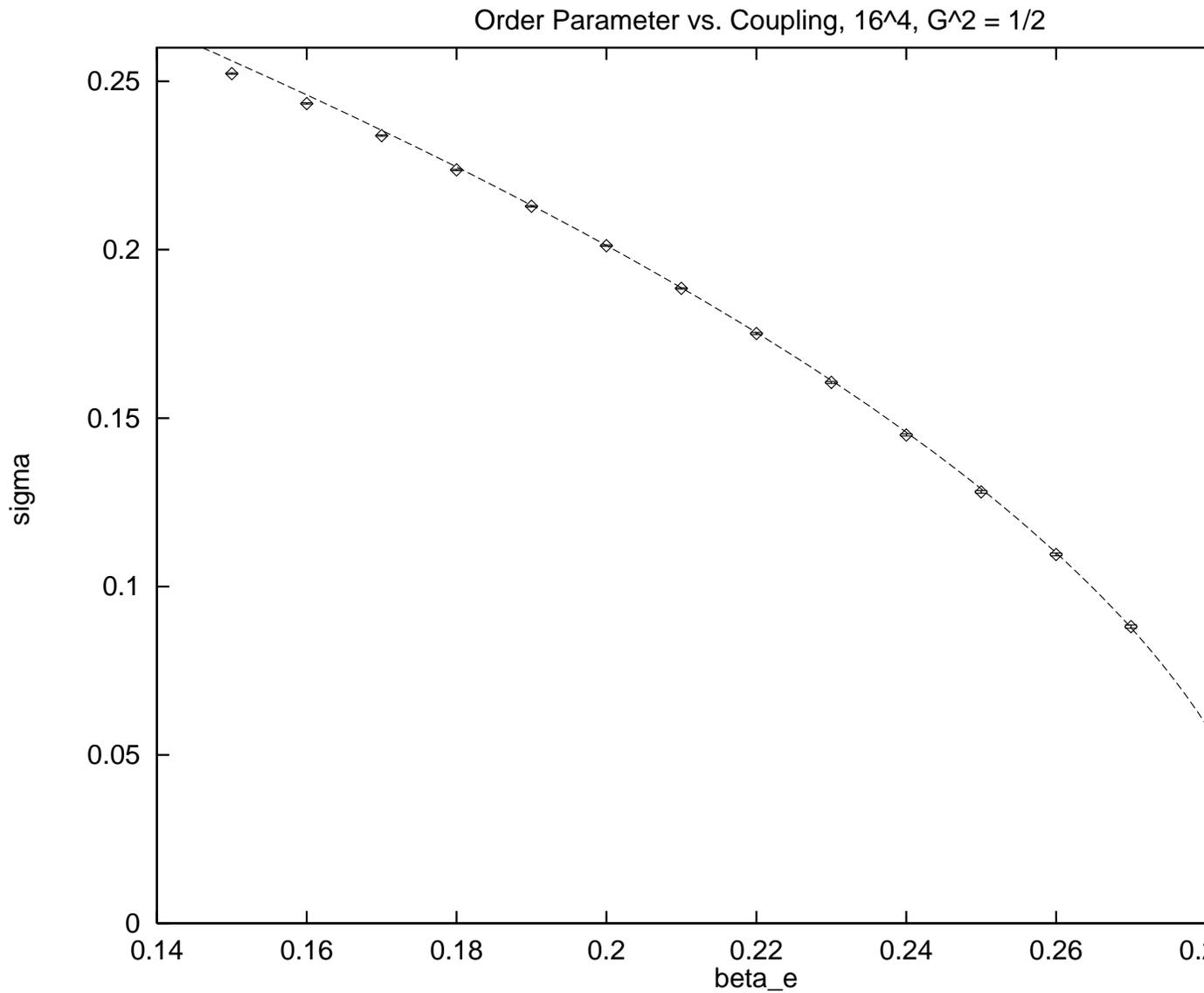,width=15cm,angle=-90,clip=}
\vspace{0.3cm}
\caption{$\sigma$ vs. $\beta_e$ for $G^2 = 1/2$}
\label{fig:sig16_z2gN4}
\end{figure}

We plot the data and the fit as shown in
Fig.7, $|\beta_c-\beta_e|/\sigma^2$ vs. $\ln(1/\sigma)$ to illustrate the importance
and numerical significance of the logartihm. 
The dashed line is the previous fit redrawn in this format.

\begin{figure}
\epsfig{file=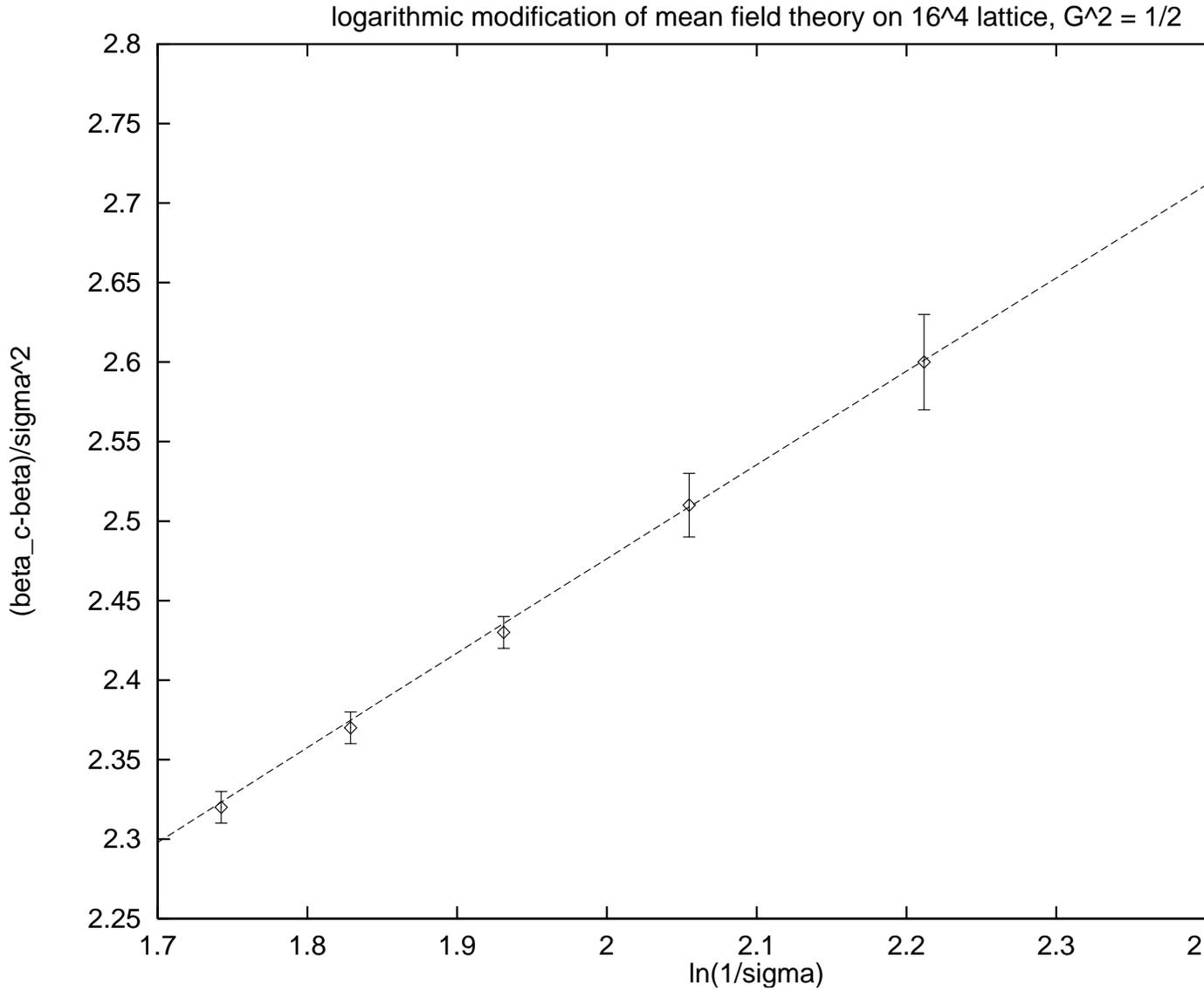,width=15cm,angle=-90,clip=}
\vspace{0.3cm}
\caption{$|\beta_c-\beta_e|/\sigma^2$ vs. $\ln(1/\sigma)$ for $G^2 = 1/2$}
\label{fig:sigz2L4}
\end{figure}

The success of this fit reiterates the irrelevance of the four fermi term : the
scaling law for the order parameter is the same as that at the  $G^2$ values of $1/8$ and $1/4$ although 
the lattice parameters, such as the location of the critical point, have changed.

Next, in Fig.8 we show the inverse of the
longitudinal susceptibility of the auxiliary field $\sigma$ 
at fixed $G^2 = 1/2$ and variable $\beta_e$.

\begin{figure}
\epsfig{file=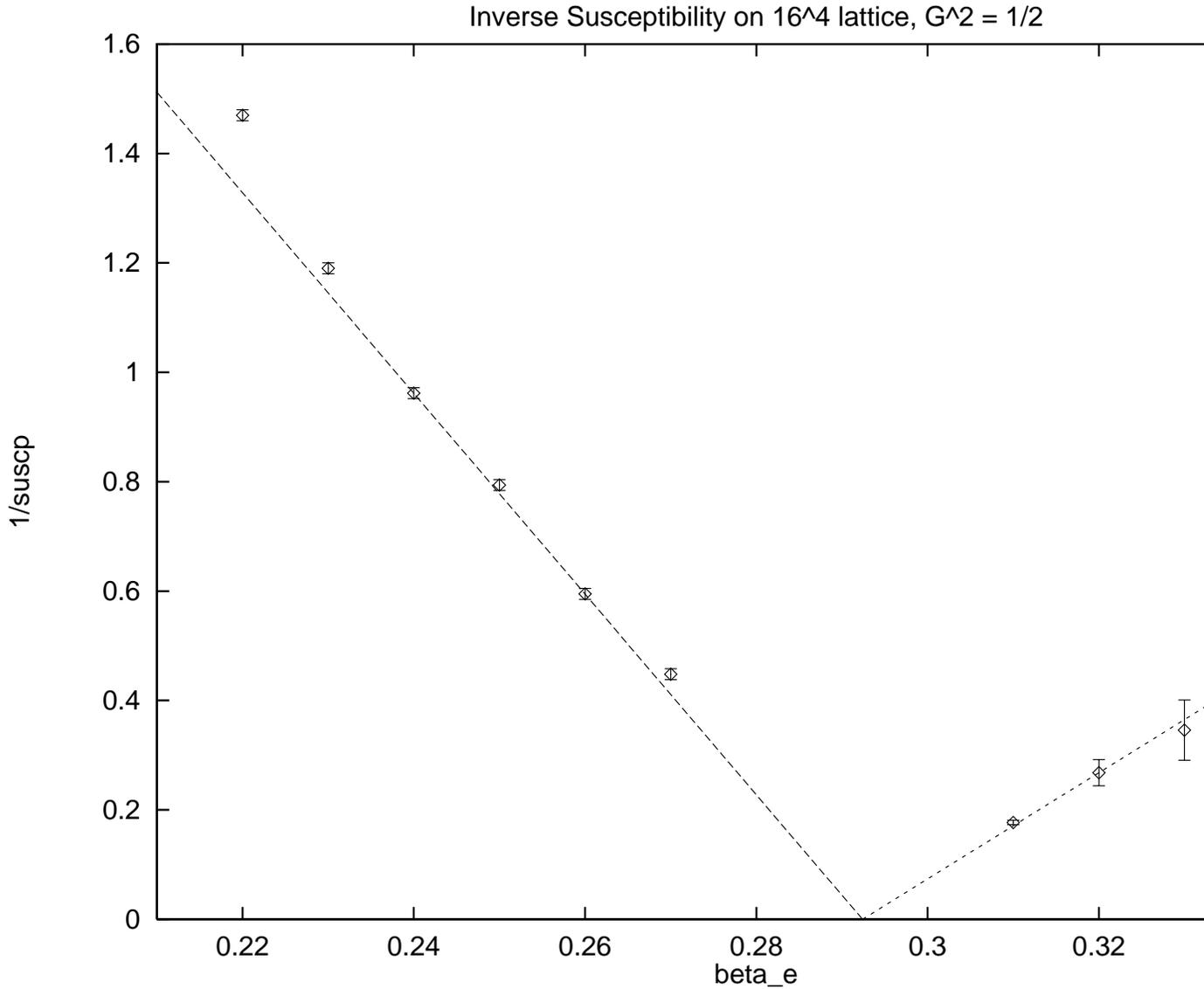,width=15cm,angle=-90,clip=}
\vspace{0.3cm}
\caption{Inverse Susceptibility vs. Coupling $\beta_e$, $G^2 = 1/2$}
\label{fig:invsusz2g}
\end{figure}

The plot picks out a critical point $\beta_c = .2924(10)$ and  
the constrained linear fits to 
the data shown in the figure produced the amplitude ratio $c_-/c_+ = 1.89(20)$. 
which compares well to the theoretical prediction
$c_-/c_+ = 1.72(2)$ . Again, the 
agreement between the simulation data and theory is good, but is not comparable in
quality or decisiveness to our order parameter fits.

\section{Simulations at nonzero Fermion Mass and the Width of the Scaling Window.}

Past simulations of lattice QED had to be done at nonzero fermion mass \cite{Dag}. The 
standard algorithms fail to converge in the limit $m \rightarrow 0$ because the
lattice Dirac operator becomes singular in the chiral limit \cite{HMC}, \cite{HMD}. This
algorithmic problem has led to indecisive results for lattice QED because of
large statistical and systematic errors. It is interesting to use the algorithm of this paper to
discover, assess and clarify the problems in past work in this field.

We chose to do simulations at nonzero fermion mass at the critical coupling $\beta_c$
determined by our fits presented in the previous section. In this way we can look for
the width of the scaling window in the $m$-direction in a particularly
simple fashion. Recall that at criticality the order parameter $\sigma$ should scale with the
fermion mass $m$, an explicit symmetry breaking parameter, as

\begin{eqnarray}
m \sim \sigma^{\delta} \ln^q(1/\sigma)
\end{eqnarray}

\noindent
where the critical index $\delta$ should be $3$ in a logarithmically trivial theory and
$q$, the power of the logarithm should be $-1$ for a $\phi^4$ theory and should be $+1$ for
a Nambu-Jona Lasinio model \cite{Koc}. By accumulating data over a range of small $m$ values, we
can look for the region where Eq. 6 might apply and determine the width of the scaling
window. It is important to keep the number of variables and parameters to a minimum in
this sort of investigation. This is the reason we work at criticality.

The critical coupling has been determined to be $\beta_c = 0.2352$ in Sec.4. The data for the order
parameter and its susceptibility are shown in Table 5 for $m$ ranging from $0.003$ to
$0.20$. Note that the statistics for this data set is particularly high as smaller and
smaller $m$ values are considered and the critical point is approached. The error bars in
$\sigma$ recorded there account for critical slowing down which forced us to accumulate
such high statistics. The statistics are at least
an order of magnitude greater than those of past studies and produce $\sigma$ values with errors ranging
from $1/2$ percent to $0.08$ percent. 

We learned in past studies of the pure Nambu-Jona Lasinio that
small $m$ values, typically below $0.01$, are needed to find a scaling window \cite{Looking}. 
However, in this case
the dynamics is controlled by the gauge coupling which alone is driving chiral symmetry breaking.
The four fermi coupling is tiny and is not affecting the dynamics in a numerically significant fashion. 
Therefore, the width of the scaling window must be determined anew from the data in Table 5.

\begin{figure}
\epsfig{file=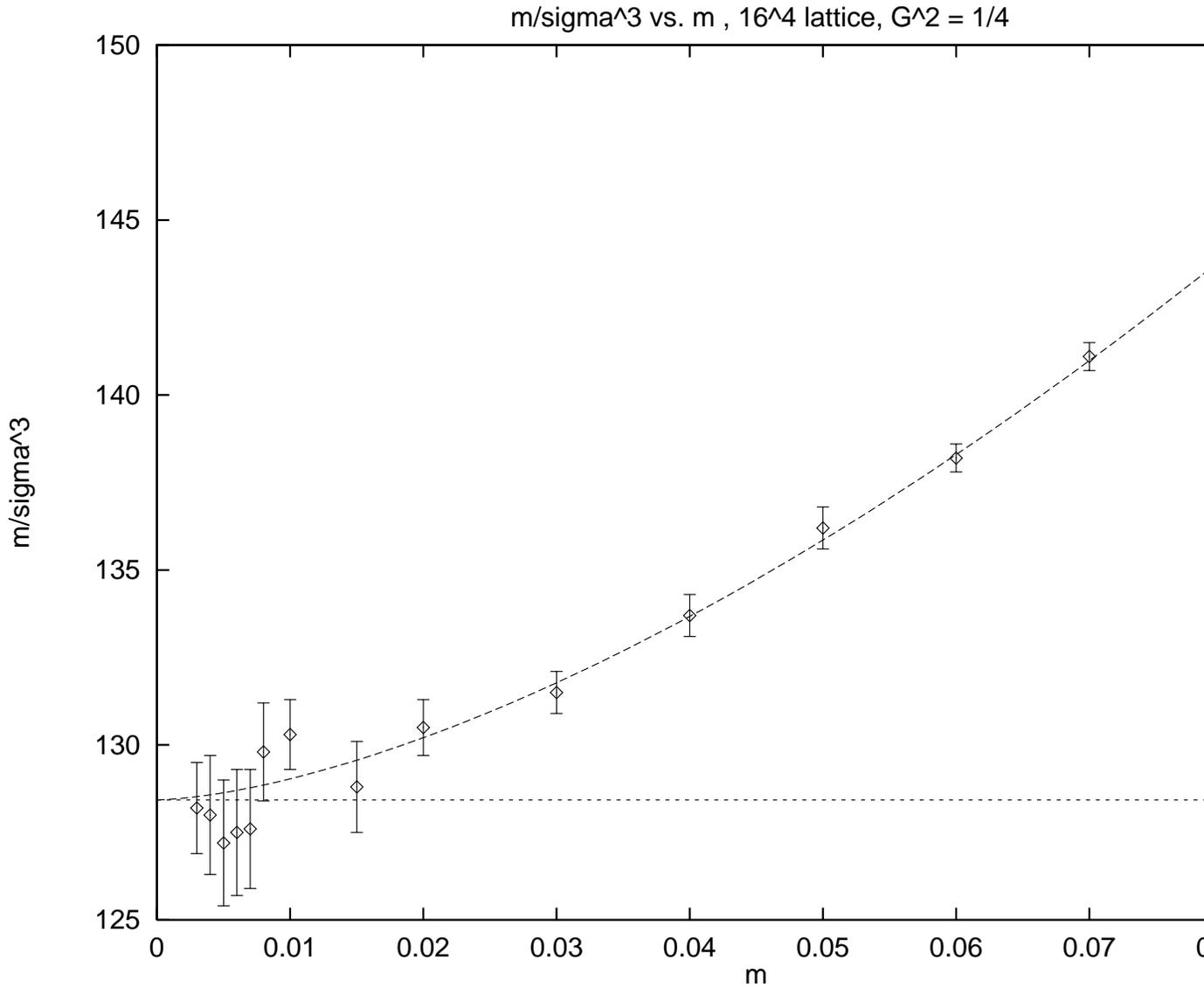,width=15cm,angle=-90,clip=}
\vspace{0.3cm}
\caption{$m/\sigma^3$ vs. $m$}
\label{fig:subdom}
\end{figure}

In Fig. 9 we plot $m/\sigma^3$ vs. $m$ in order to assess visually the relevance of the
leading logarithm result Eq. 6. The data clearly pick out the value $\delta = 3.0$ for the
dominant powerlaw singularity of the scaling law for very small $m$ values, all less than $0.01$. However,
we also see that the deviations from the mean field result are numerically significant over
the entire mass range shown. In fact, they are far too large to be accommodated by a weak 
logarithmic scaling violation as expected in Eq. 6. In fact, a fit of that form to the data
ranging from $m = 0.003$ to $m = 0.08$ produces a huge value for the power of the logarithm, $q = -8(1)$,
and a very small confidence level of $0.43$ percent ($\chi^2/d.o.f. \approx 28.8/12$). Therefore,
the data rule out the applicability of logarithmic improved mean field scaling to describe
the data at nonzero $m$ except for the very smallest values of $m$, $m < 0.01$. Unfortunately, most
data used to study the potential triviality of QED using the conventional action  
employed $m$ values considerably larger than $m = 0.01$ in order
to run efficiently and generate sufficient statistics. Typical ranges of $m$ have been between
$0.01$ and $0.10$ \cite{Goc} and are very sensitive to data taken with $m = 0.02$, $0.03$ and $0.04$.
This criticism applies to all past studies of noncompact QED, for example, \cite{Kogut,KW,Dag}. It
also means that the methods of analysis introduced in \cite{referee8} do not apply to this data
set because those methods require data in a scaling window, controlled by a single asymptotic form.
Higher precision data taken at the smallest values of $m$, $m < 0.01$, are required apparently and, in
fact, larger lattices than $16^4$ might be necessary also because of the possibility of significant
finite size effects.

Two possible explanations for the data come to mind : 1. Perhaps the real critical point 
is significantly different
from $0.2352$ as determined by our fits at $m = 0.0$, or 2. Perhaps subdominant singularities in
the scaling law are numerically significant over this range of $m$. 

It is easy to rule out option 1. Ignoring logarithms, the mean field equation of state reads
$m = D \sigma^3 -C (\beta_c-\beta_e) \sigma $. This implies that if $\beta_e$ were
different from $\beta_c$, then $m/\sigma^3$ would behave as $D - C (\beta_c-\beta_e)/ \sigma^2$,
and the correction term would be large for small $\sigma$, which is just the opposite of the
behavior observed in Fig. 9.

Now consider option 2. If a subdominant singularity contributes to the equation of state, then
at criticality the relation $B \sigma^3 = m$ should be replaced by,

\begin{eqnarray}
m = B \sigma^{\delta} + D \sigma^{\delta_s}
\end{eqnarray}

\noindent
where $\delta$ should be $3$ and $\delta_s$ should be considerably larger \cite{Az}. This hypothesis fits
the data beautifully : the curved dashed line in Fig. 9 shows the fit which has a confidence level
of $98.7$ percent ($\chi^2/d.o.f. \approx 3.78/12$). In fact this fitting form can be well approximated
in a fashion that is useful for practical purposes,

\begin{eqnarray}
m/\sigma^3 \approx B + D^{\prime} m^{\delta_s/\delta}
\end{eqnarray}

\noindent
Eq. 8 approximates Eq. 7 because the correction to the constancy of $m/\sigma^3$ is
less than $15$ percent over the range of
$m$ values in the figure. The fit gives $B=128.43(58)$, $D^{\prime} = 795(311)$, and
$\delta_s  = 4.56(16)$.

We learn several lessons from this exercise.

Previous simulations of pure QED at nonzero $m$ could not possibly have detected the
logarithms of triviality decorating mean field singularities. For the present range of $m$ values and
lattice sizes, data at nonzero $m$ have contributions from subdominant critical singularities
which are larger numerically than logarithmic corrections to mean field theory.

\section{Finite Size Effects}

Since we are using a new algorithm which works in the limit of massless quarks, we
should be careful to monitor finite size effects. Some of our data are taken very near to
critical points in order to find critical indices that control continuum limits
of the lattice models. At these points the model's correlation length diverges and
there are potentially dangerous finite size effects which could mimic  finite temperature
effects, for example. We need to check that the lattice is large enough to contain
correlations larger than the lattice spacing but smaller than the system's spatial extent
in order to work within a scaling window where we can
extract continuum features of the field theory.

In Table 6 we show data for $\sigma$ taken for gauge couplings $\beta_e$ ranging from $0.15$
to $0.27$ at fixed four fermi coupling $G^2=1/2$ for $12^4$, $16^4$ and $20^4$ lattices. The comparison
of the three data sets shows coincidence everywhere except at $\beta_e = 0.27$ between the smallest
lattice $12^4$ and the other two. $\beta_e = 0.27$ was our closest approach to the critical
point in the symmetry broken phase and it appears that our $16^4$ lattice was sufficient,
given our $1/2$ percent statistical errors. Reliance on a $12^4$ lattice would have failed us.

In the next table we show $12^4$ data for a simulation where the four fermi coupling is
fixed at $G^2 = 1/4$ and $\beta_e$ ranges from $0.15$ to $0.25$. The data consists of $\sigma$
as well as monopole observables that will be discussed in a later section. Comparing the
$\sigma$ data here to that in Table 3, we confirm the absence of finite size effects within
our statistical errors.

In summary, the $16^4$ data we have used to extract scaling laws from $\sigma$ measurements
appears free of significant finite size effects. The significance of finite size effects depends
strongly on the observable being simulated. We also checked that the longitudinal susceptibility
data that was used to extract the logarithmic violations of scaling in the
amplitudes was not distorted by finite size effects. Since these susceptibilities are
determined with much larger statistical error bars, this test was less demanding. Certainly
the finite size effects in $\chi_{\sigma}$ are much larger than those in $\sigma$ itself. However,
since $\sigma$ was determined within a fraction of a percent while the statistical uncertainty in
$\chi_{\sigma}$ was typically several percent, a $16^4$ lattice was adequate for the range of
couplings used in this study.

\section{Monopole Observables}

Noncompact lattice QED was first studied with the goal of simulating the dynamics of $U(1)$
gauge fields without the monopoles that accompany compact lattice QED \cite{Banks}. It was found, however,
that even the noncompact formulation has monopole-like dislocations in its lattice formulation
because of the space-time cutoff itself \cite{HW}. These dislocations can undergo a percolation transition
where long range correlations develop \cite{HW}. Because of this transition, it is not obvious that
simulation results in pure noncompact lattice QED reflect the physics of textook QED in which
field configurations are smooth and have no topological excitations.
The formulation of noncompact lattice QED with a four fermi term is free
of the issues raised in \cite{HW}. The point is, as discussed in the Section 2 above, the 
monopole percolation transition does not coincide with the chiral transition as long as the four 
fermi coupling is nonzero. Therefore the gauge field vacuum is free of critical dislocations
at the gauge couplings of interest, so we know that we are studying a model free of
topological excitations, as we wish.

Let's find the monopole percolation transition in the model with a fixed four fermi coupling
$G^2 = 1/4$. The data for the monopole concentration $M$ and the associated monopole
percolation susceptibility $\chi_M$, both defined exactly
as in \cite{HW}, are given in Table 2. In Fig.10 we plot the monopole concentration
against the gauge coupling and find a percolation transition at $\beta_c^M = .2104(1)$.

\begin{figure}
\epsfig{file=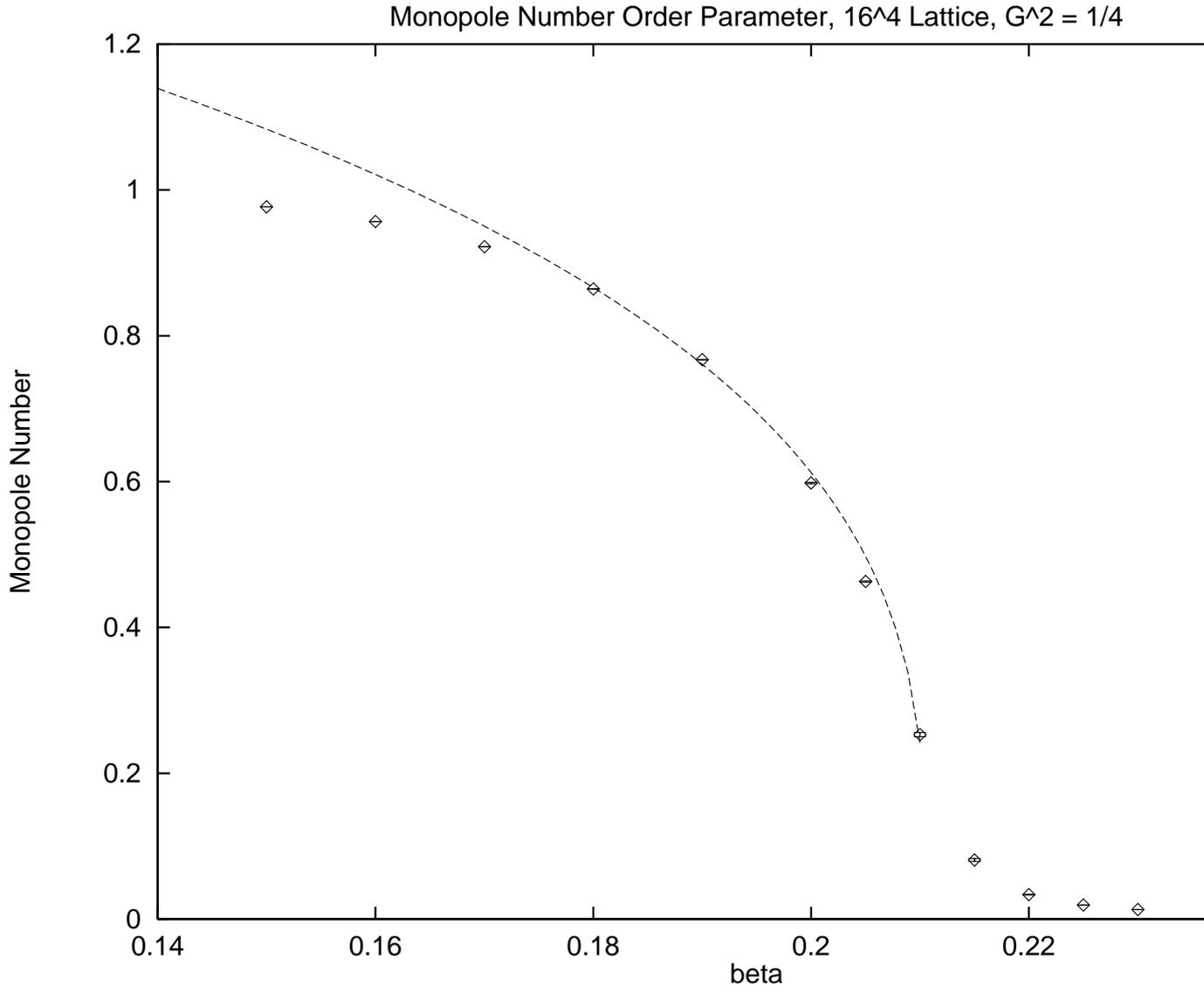,width=15cm,angle=-90,clip=}
\vspace{0.3cm}
\caption{Monopole Concentration $M$ vs. Coupling $\beta_e$, $G^2 = 1/4$}
\label{fig:mon16}
\end{figure}

We determined in Sec.4 that the chiral transition occurs at considerably weaker 
coupling, $\beta_c = .2352$, where the monopole
concentration is insignificant, as we read off Fig.10.

It is also informative to confirm this conclusion by considering the monopole percolation
susceptility, $\chi_M$. In Fig.11 we plot
this susceptibility against the gauge coupling and see that it
appears to diverge in the vicinity of $\beta_c^M = .2104(1)$ (we also confirmed this impression
with powerlaw fits). In addition, in Fig.12 we plot the longitudinal susceptibility of the chiral transition
and confirm that it diverges near $\beta_c = .2352$, as already determined in Sec.4. 
The two susceptibility peaks are cleanly separated : $\beta_c^M = .2104(1)$ vs. $\beta_c = .2352$.

\begin{figure}
\epsfig{file=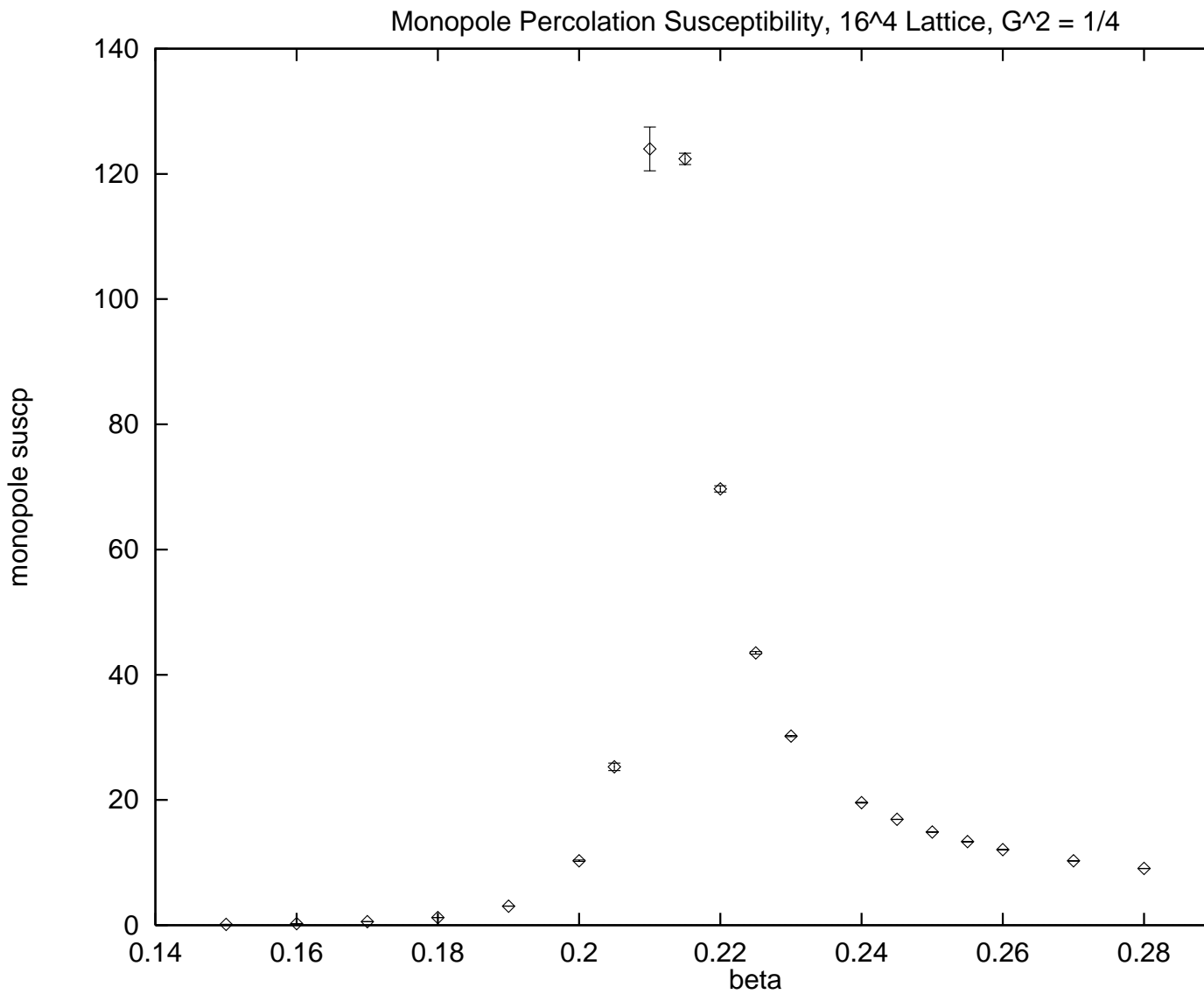,width=15cm,angle=-90,clip=}
\vspace{0.3cm}
\caption{Monopole Percolation Susceptibility $M$ vs. Coupling $\beta_e$, $G^2 = 1/4$}
\label{fig:susmon16}
\end{figure}

\begin{figure}
\epsfig{file=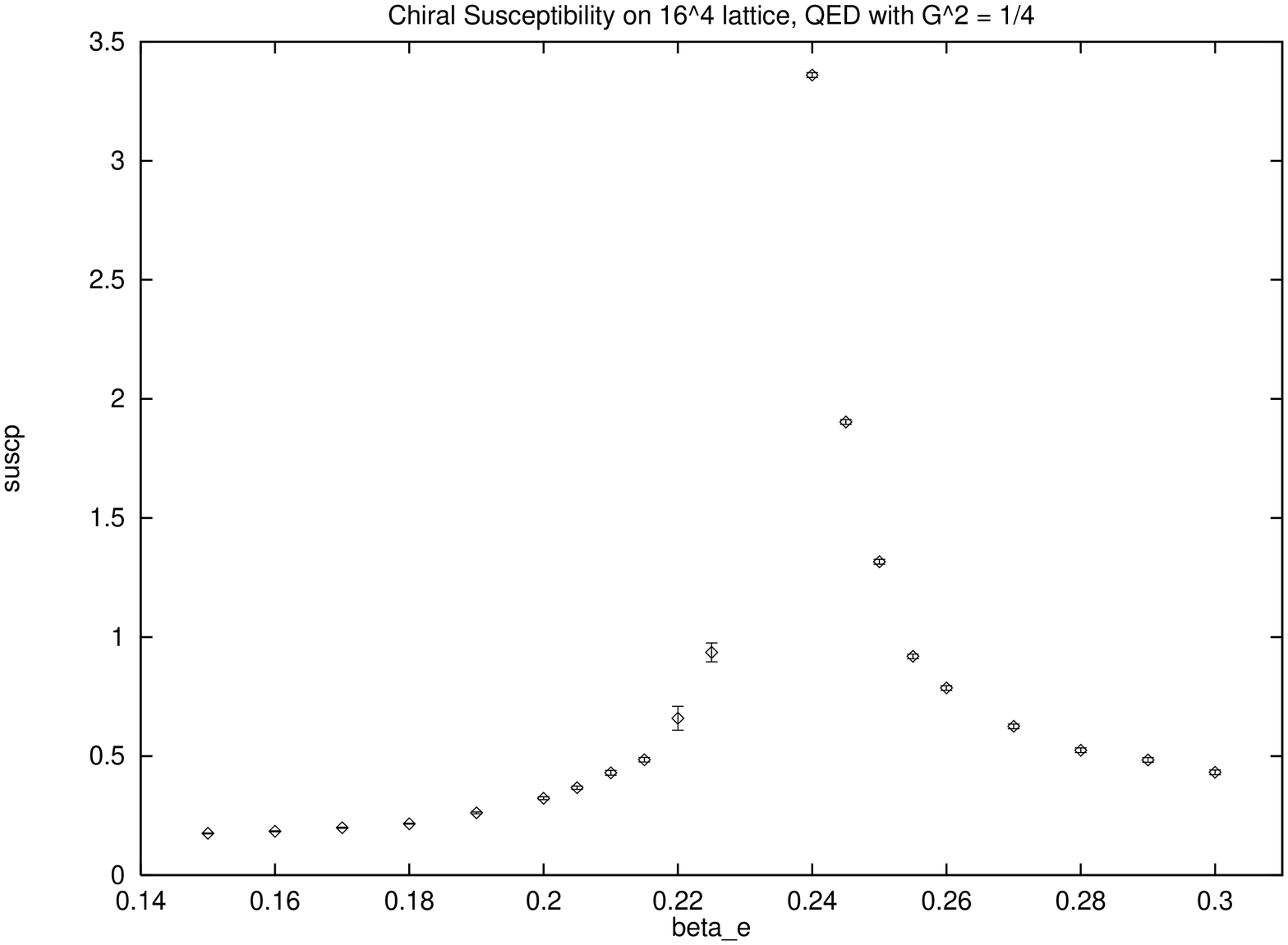,width=15cm,angle=-90,clip=}
\vspace{0.3cm}
\caption{Longitudinal Chiral Susceptibility $M$ vs. Coupling $\beta_e$, $G^2 = 1/4$}
\label{fig:sus168}
\end{figure}

We end this section with a minor remark about the finite size effects observed in the monopole observables.
Comparing Table 2 and Table 7, we see that as the monopole percolation transition's critical coupling
is approached, there are numerically significant differences between the $12^4$ and the $16^4$ data sets
for both the concentration $M$ as well as its associated susceptibility $\chi_M$. As expected, the
percolation susceptibility $\chi_M$ is strongly suppressed by the lattice size near the transition. In fact,
as we have discussed elsewhere \cite{HKK}, finite size scaling of the peak of the susceptibility is
an effective and accurate means to measure the percolation critical indices. It would take simulations
on a series of lattice sizes to carry out such a program for this model. The only point we wish to make here,
however, is that the percolation and chiral transitions are well separated inside the
phase diagram Fig. 1. It is interesting (and fortunate for the success of this project) that
the finite size effects in the chiral order parameter $\sigma$ are
significantly smaller than those in the monopole concentration.

\section{Failures and Challenges at $G^2=0$}

Although the major topic in this research is the behavior of the gauged Nambu-Jona Lasinio
model for $G^2 \neq 0$, we will briefly discuss the present confusing state of theory and
simulations at the edge of the phase diagram $G^2=0$ where past simulations have been carried out.
As we have already emphasized, the real problem with studies at $G^2=0$ is that they must be 
done at nonzero fermion mass away from the chiral limit and this has caused several problems : 1. The
simulations become excessively slow for small $m$ values because the lattice Dirac 
operator is singular in that limit. Therefore, at low values of $m$ where the best statistics
are required, the statistics of the data sets are typically the poorest. 2. The scaling window
in the $m$-direction is extremely narrow, so fitting forms which only account for the leading
critical behavior are inadequate and misleading. Attempting to go beyond leading order
critical singularities in fits leads to a vast proliferation of parameters which undermines
firm conclusions.

Another potential problem concerning the $G^2=0$ edge of the phase diagram concerns lattice
monopoles. Recall that one motivation for inventing and studying noncompact lattice QED \cite{Shenker}
was to make a model free of monopoles in order to understand the relation between
chiral symmetry breaking and single gluon exchange. However, Hands and Wensley \cite{HW} 
pointed out that even the
the noncompact model has monopole-like lattice dislocations because of gauge invariance of
the pure gauge field piece of the action and because of the lattice cutoff itself. These authors
also pointed out that these lattice monopoles experience a percolation transition as the
gauge coupling becomes strong and in the case of quenched simulations, the monopole percolation
transition is very close to the chiral transition experienced by light fermions \cite{HW}. This led these
authors to speculate that noncompact lattice QED might not be a sound framework for studying
"textbook" QED at strong coupling \cite{HW}.

What does this possibility mean for this paper? Since we work at $G^2 \neq 0$ where the critical line of
monopole percolation is distinct from the chiral transition line, these lattice artifacts are not
relevant to our conclusions. We believe that we have a firm theoretical and numerical grasp of gauged
Nambu-Jona Lasinio models everywhere within the phase diagram Fig.1 but not along the edge
$G^2=0$. How could this be? Following Hands and Wensley, the gauge field piece of the
action Eq.2 is invariant under local gauge transformations defined by the group
of real numbers $R$, while the fermionic piece of
the action, which describes the gauge invariant hopping of the fermion around the lattice,
has a gauge symmetry based on phases, $U(1)$. The cutoff theory described by
the pure gauge piece of the action has monopole excitations attached by Dirac 
strings \cite{HW}. These are singular
field configurations whose actions diverge when the lattice spacing is taken to zero. They
would be of no concern if it weren't for the fact that as the coupling increases they 
experience a percolation transition where monopole clusters develop macroscopic dimensions.
Since the fermions are sensitive to monopole clusters through their $U(1)$ phase, Hands and
Wensley speculated that they could affect the chiral transition in the quenched and unquenched
model. This speculation could be wrong for several reasons : 1. The underlying gauge action is just 
a quadratic form, so it is a perfectly solvable free field theory. A free field theory can't have a phase
transition, as emphasized in \cite{referee12} and 2. Percolation transitions
need not affect the bulk properties of the underlying field theory.
Many examples of this sort can be cited. These complaints can be answered in part : 1. The phase transition
of percolation is not in local observables constructed out of the gauge fields, but rather is in
nonlocal matrix elements. It is not unusual in statistical mechanics to make models where non-local matrix
elements experience phase transitions when the underlying local field theory has no transition itself.
Condensed matter physics provides many examples of enormous practical importance
including, for example, the localization-delocalization transition of single electrons in
background fields of varying degrees of disorder.
The chiral transition is sensitive to loops of the $U(1)$ phase and is of this type.
2. Since fermions flip their chirality in the
presence of monopoles, it is plausible that a percolating network of monopole-like excitations
can induce chiral symmetry breaking in the bulk system. There is a possibility that the $G^2 = 0$ pure
QED model has qualitatively different physics than that found anywhere within the phase diagram
in Fig.1. Only at the edge of the diagram would the percolating monopole-like excitations be
critical where chiral symmetry is broken. Only there are new degrees of freedom, percolating
monopoles, relevant so only there could there be a new universality class. It might be that
on the edge of the phase diagram, the chiral condensate is driven by monopole percolation and
the chiral transition inherits a correlation length critical index $\nu \approx 2/3$ from
the percolating network and becomes the basis for a nontrivial field theory \cite{HKKperc}.
It has been noticed that as the number of fermions is varied, both the chiral and monopole
percolation transitions move in unison \cite{KW}.
In addition, in unquenched models, such as the
four flavor model on the edge of the phase diagram Fig.1, the fermions induce $U(1)$ plaquette
terms into the theory's action which can support conventional lattice monopoles \cite{Banks}.

We have nothing to add to the pros and cons of these qualitative arguments. We hope that
the physics issues brought up here could be answered by striking out in new directions and
finding approaches or arguments which are more precise and quantitative. The monopole percolation
picture may contain only half truths, but some of those ideas might be testable in the context
of models with real monopoles, generalizations of compact $U(1)$ lattice QED \cite{Banks}, perhaps.

\section {Conclusions and discussion}

We presented numerical evidence for the triviality of textbook QED using a new algorithm which converges
for massless quarks. Past simulations using the action with massive quarks but no four fermi term
produced controversial results.
Recall that \cite{Kogut,HKKperc} claimed non-triviality for the theory while
\cite{Goc,referee4} found triviality and backed up their claim further in
\cite{Goc} by calculating the sign of the beta function, which is directly relevant
to the question of triviality.

It would be worthwhile to continue using the new algorithm and pursue several new directions.

One could calculate the theory's renormalized couplings and their RG trajectories in the
chiral limit, extending the work of \cite{Goc2} to a two parameter space. Both the gauge and the Four Fermi
couplings should vanish as the reciprocal of the logarithm of the ultra-violet cutoff. As discussed in \cite{Goc2}
this calculation has some technical challenges specific to lattices of finite extent which necessitate
the extrapolation of raw lattice data to achieve physical results. It would be worthwhile to
investigate improved strategies here to avoid crude, indecisive results. The high quality of the equation of
state fits in Sec.4, 5 and 6 should lead to improved determinations of the renormalized couplings
because the lattice critical couplings are determined with excellent precision.

One could also simulate the model 
with the $Z_2$ chiral group replaced by a continuous group 
so the model would have Goldstone bosons even on a coarse lattice \cite{Az3}. It
would then be possible to test the approach and results of  \cite{Az3} more quantitatively.

It would also be interesting to generalize the results of Sec.7, that a subdominant critical singularity
is needed to describe the data at nonzero $m$, away from the critical coupling. In other words,
fit the finite $m$ data points of previous investigations such as \cite{Goc,referee4} to equations of state with
both a dominant and subdominant singularity and check that improved confidence levels are achieved with simple
hypotheses. Unfortunately, there will be a proliferation of fitting parameters in such a program, so
its numerical significance might be questioned. Nonetheless, it would definitely be worth consideration. Such
a program would also influence the determination of renormalized couplings because these calculations
use critical couplings inferred from equation of state fits \cite{Goc2}.

Finally, it would be 
interesting to simulate 
compact QED with a small four fermi term and study the interplay of monopoles, charges and
chiral symmetry breaking. Since the $G = 0$ limit of the compact model 
is known to have a first order transition \cite{EDK},
generalizations of the action will be needed to
find a continuous transition where a continuum limit of the lattice theory might exist. Since
the parameter space of the generalized model would be at least three dimensional, this
interesting problem would be quite challenging.

\centerline{Acknowledgment}

This work was partially supported by NSF under grant NSF-PHY96-05199. S. K. is
supported by the Korea Research Foundation. M.-P. L. wishes to thank the 
{\em ECT$^*$}, Trento, for hospitality during the final stages of this 
project.  The simulations were done at NPACI and NERSC. 


\mediumtext

\begin{table}
\begin{tabular} {cdd} 
$dt$ & $\sigma$ &    $Trajectories$   \\ \hline
$.01$ &   0.1279(6)     & 1200  \\
$.02$ &   0.1280(5)     & 2000  \\
$.03$ &   0.1284(4)     & 2400  \\
$.04$ &   0.1293(4)     & 2800  \\
$.05$ &   0.1301(3)     & 3500  \\
$.06$ &   0.1314(2)     & 4800  \\
\end{tabular}
\caption{ Dependence of $\sigma$ on $dt$ for a $12^4$ lattice with four fermi coupling 
$\lambda= 4.0$ and gauge coupling $\beta=0.25$.}
\end{table}
\narrowtext

\begin{table}
\begin{tabular} {cddddd} 
$\beta_g$ & $\sigma$ &  $\chi_{\sigma}$   &  $M$   &  $\chi_M$   &  $Trajectories$   \\ \hline
$.150$ &   0.11980(7) &  0.1756(5)    &  0.97676(1)   &  0.1245(5)   & 930  \\
$.160$ &   0.11248(8) &  0.1843(20)   &  0.9566(1)    &  0.261(1)    & 950  \\
$.170$ &   0.10438(8) &  0.1993(50)   &  0.9221(1)    &  0.552(2)    & 1030  \\
$.180$ &   0.09531(9) &  0.2160(50)   &  0.8644(2)    &  1.220(5)    & 1010  \\
$.190$ &   0.08520(8) &  0.2621(30)   &  0.7669(3)    &  3.04(2)     & 1500  \\
$.200$ &   0.0738(1)  &  0.3230(40)   &  0.5976(7)    &  10.3(1)     & 1310  \\
$.205$ &   0.0674(1)  &  0.3671(30)   &  0.463(1)     &  25.3(6)     & 1012  \\
$.210$ &   0.0609(1)  &  0.4301(30)   &  0.250(2)     &  124.0(9)    & 1130  \\
$.215$ &   0.0537(2)  &  0.4849(40)   &  0.0812(8)    &  122.4(9)    & 2701  \\
$.220$ &   0.0456(2)  &  0.6591(40)   &  0.0338(5)    &  69.7(5)     & 1120  \\
$.225$ &   0.0367(2)  &  0.9356(40)   &  0.0192(2)    &  43.5(2)     & 1670  \\
$.230$ &              &               &  0.0130(2)    &  30.2(1)     & 810  \\
$.240$ &  -0.00008(9) &  3.360(90)    &  0.00798(8)   &  19.57(5)    & 810  \\
$.245$ &   0.0002(5)  &  1.903(80)    &  0.00681(4)   &  16.90(2)    & 2518  \\
$.250$ &  -0.00002(3) &  1.316(70)    &  0.00601(5)   &  14.87(3)    & 960  \\
$.255$ &   0.0003(4)  &  0.919(50)    &  0.00544(2)   &  13.34(1)    & 3350  \\
$.260$ &  -0.0007(3)  &  0.786(40)    &  0.00483(4)   &  12.07(2)    & 820  \\
$.270$ &  -0.0006(3)  &  0.625(50)    &  0.00407(4)   &  10.28(1)    & 1010  \\
$.280$ &  -0.0003(2)  &  0.525(10)    &  0.00303(6)   &  9.051(9)    & 1070  \\
$.290$ &  -0.0002(2)  &  0.484(10)    &  0.00166(6)   &  8.185(7)    & 1230  \\
$.300$ &   0.0002(2)  &  0.432(10)    &  0.00074(5)   &  7.521(6)    & 1150  \\
\end{tabular}
\caption{ Observables measured on a $16^4$ lattice with four fermi coupling 
$G^2 = 1/4$}
\end{table}
\narrowtext

\begin{table}
\begin{tabular} {cddddd}
$\beta_g$ & $\sigma$ &  $\chi_{\sigma}$   &  $M$   &  $\chi_M$   &  $Trajectories$   \\ \hline
$.160$ &   0.05201(4)  &  0.066(2)   &  0.9479(1)     &  0.327(1)   & 730  \\
$.165$ &   0.04932(4)  &  0.068(1)   &  0.9293(1)     &  0.487(2)   & 770  \\
$.170$ &   0.04642(5)  &  0.075(2)   &  0.9047(2)     &  0.729(3)   & 660  \\
$.175$ &   0.04332(5)  &  0.074(2)   &  0.8729(3)     &  1.106(5)   & 600  \\
$.180$ &   0.03996(6)  &  0.082(2)   &  0.8314(4)     &  1.72(1)    & 500  \\
$.185$ &   0.03644(6)  &  0.089(3)   &  0.7765(5)     &  2.81(2)    & 640  \\
$.190$ &   0.03267(5)  &  0.096(3)   &  0.7054(6)     &  4.82(3)    & 960  \\
$.195$ &   0.02848(7)  &  0.110(4)   &  0.6091(9)     &  9.35(9)    & 770  \\
$.200$ &   0.02401(9)  &  0.136(4)   &  0.4751(16)    &  24.2(9)    & 640  \\
$.205$ &   0.01872(9)  &  0.190(5)   &  0.2677(29)    &  115(4)     & 780  \\
$.210$ &   0.01250(9)  &  0.38(1)    &   0.08801(19)  &  127(2)     & 960  \\
$.215$ &  -0.00013(64) & 1.62(9)     &   0.03625(95)  &  74.1(9)    & 310  \\
$.220$ &  -0.00027(38) & 0.75(9)     &    0.02333(52) &  51.3(7)    & 320  \\
$.225$ &   0.00009(15) & 0.33(4)     &   0.01576(26)  &  35.6(2)    & 490  \\
$.230$ &   0.00009(9)  &  0.23(4)    &   0.01172(15)  &  27.9(1)    & 660  \\
$.235$ &   0.00009(9)  &  0.19(4)    &   0.00937(12)  &  22.8(8)    & 590  \\
$.240$ &   0.00010(8)  &  0.16(2)    &   0.00779(9)   &  19.22(6)   & 640  \\
$.245$ &   0.00010(6)  & 0.145(9)    &  0.00665(6)    &  16.63(4)   & 910  \\
$.250$ &   0.00009(6)  &  0.133(7)   &  0.00586(5)    &  14.67(3)   & 790  \\
\end{tabular}
\caption{ Observables measured on a $16^4$ lattice with four fermi coupling
$G^2 = 1/8$}
\end{table}
\narrowtext

\begin{table}
\begin{tabular} {cddd} 
$\beta_g$ & $\sigma$  &  $\chi_{\sigma}$      &  $Trajectories$   \\ \hline
$.150$ &   0.2525(2)  &  0.39(5)      & 1500  \\
$.160$ &   0.2434(2)  &  0.43(2)      & 1500  \\
$.170$ &   0.2339(2)  &  0.43(5)      & 1500  \\
$.180$ &   0.2237(2)  &  0.44(5)      & 1500  \\
$.190$ &   0.2129(2)  &  0.48(3)      & 1500  \\
$.200$ &   0.2012(2)  &  0.51(4)      & 1500  \\
$.210$ &   0.1885(2)  &  0.61(3)      & 1500  \\
$.220$ &   0.1751(3)  &  0.68(3)      & 1500  \\
$.230$ &   0.1606(3)  &  0.84(2)      & 1500  \\
$.240$ &   0.1450(3)  &  1.04(5)      & 1500  \\
$.250$ &   0.1281(4)  &  1.26(4)      & 1500  \\
$.260$ &   0.1095(4)  &  1.68(4)      & 1500  \\
$.270$ &   0.0881(5)  &  2.23(5)      & 1500  \\
$.310$ &   0.000(6)   &  5.66(10)     & 1500  \\
$.320$ &   0.000(7)   &  3.68(10)     & 1500  \\
$.330$ &   0.000(7)   &  3.17(10)     & 1500  \\
$.340$ &   0.000(6)   &  2.61(10)     & 1500  \\
$.350$ &   0.000(4)   &  2.33(10)     & 1500  \\
$.360$ &   0.000(5)   &  2.31(7)      & 1500  \\
$.370$ &   0.000(5)   &  1.80(6)      & 1500  \\

\end{tabular}
\caption{ Observables measured on a $16^4$ lattice with four fermi coupling 
$G^2 = 1/2$}
\end{table}
\narrowtext

\begin{table}
\begin{tabular} {cddd} 
$m$    & $\sigma$ &  $\chi_{\sigma}$     &  $Trajectories$   \\ \hline
$.003$ &   0.0286(1)  &  0.927(9)       & 3400  \\
$.004$ &   0.0315(1)  &  0.779(2)       & 3194  \\
$.005$ &   0.0340(1)  &  0.681(4)       & 3040  \\
$.006$ &   0.0361(1)  &  0.639(5)       & 3397  \\
$.007$ &   0.0380(1)  &  0.582(8)       & 3207  \\
$.008$ &   0.03944(9) &  0.538(5)       & 2174  \\
$.010$ &   0.04247(9) &  0.474(5)       & 1837  \\
$.015$ &   0.04890(8) &  0.415(5)       & 2521  \\
$.020$ &   0.05357(8) &  0.377(4)       & 1740  \\
$.030$ &   0.06110(9) &  0.325(8)       & 1260  \\
$.040$ &   0.06688(8) &  0.290(7)       & 1380  \\
$.050$ &   0.07160(7) &  0.269(7)       & 1400  \\
$.060$ &   0.07571(8) &  0.255(6)       & 960  \\
$.070$ &   0.07916(7) &  0.240(6)       & 1020  \\
$.080$ &   0.08214(8) &  0.231(7)       & 1000  \\
$.090$ &   0.08480(8) &  0.217(6)       & 910  \\
$.100$ &   0.08724(7) &  0.211(6)       & 970  \\
$.150$ &   0.09636(7) &  0.193(3)       & 970  \\
$.200$ &   0.10237(8) &  0.177(3)       & 730  \\
\end{tabular}
\caption{ Criticality runs on a $16^4$ lattice with four fermi coupling 
$G^2 = 1/4$ and variable fermion mass $m$}
\end{table}
\narrowtext

\begin{table}
\begin{tabular} {cddd} 
$\beta_g$ & $\sigma, 12^4$  &  $\sigma, 16^4$   &  $\sigma, 20^4$    \\ \hline
$.150$ &            &   0.2525(2) &                \\
$.160$ &            &   0.2434(2) &                \\
$.170$ &  0.2341(4) &   0.2339(2) &                \\
$.180$ &  0.2239(4) &   0.2237(2) &                \\
$.190$ &  0.2130(4) &   0.2129(2) &                \\
$.200$ &  0.2013(5) &   0.2012(2) &                \\
$.210$ &  0.1885(5) &   0.1885(2) &                \\
$.220$ &  0.1747(6) &   0.1751(3) &  0.1748(4)     \\
$.230$ &  0.1606(6) &   0.1606(3) &  0.1617(5)     \\
$.240$ &  0.1451(7) &   0.1450(3) &  0.1454(3)     \\
$.250$ &  0.1281(6) &   0.1281(4) &  0.1283(4)     \\
$.260$ &  0.1089(8) &   0.1095(4) &  0.1093(4)     \\
$.270$ &  0.0866(8) &   0.0881(5) &  0.0885(4)     \\

\end{tabular}
\caption{ Chiral condensate $\sigma$ on $12^4$, $16^4$, and $20^4$ lattices with 
four fermi coupling $G^2 = 1/2$. Finite size study.}
\end{table}
\narrowtext

\begin{table}
\begin{tabular} {cdddd} 
$\beta_g$ & $\sigma$      &  $M$   &  $\chi_M$   &  $Trajectories$   \\ \hline
$.150$ &   0.1203(2)     &  0.9770(1)    &  0.123(1)    & 1000  \\
$.160$ &   0.1129(2)     &  0.9568(2)    &  0.261(2)    & 1000  \\
$.170$ &   0.1047(2)     &  0.9223(4)    &  0.551(5)    & 1000  \\
$.180$ &   0.0956(2)     &  0.8641(6)    &  1.21(1)     & 1000  \\
$.190$ &   0.0854(3)     &  0.7660(10)   &  3.10(5)     & 1000  \\
$.200$ &   0.0741(3)     &  0.5954(22)   &  10.6(3)     & 1000  \\
$.210$ &   0.0610(4)     &  0.2669(49)   &  71.5(9)     & 1000  \\
$.220$ &   0.0449(7)     &  0.0651(16)   &  51.0(7)     & 1000  \\
$.230$ &                 &  0.0291(5)    &  26.6(2)     & 1000  \\
$.240$ &                 &  0.0203(6)    &  18.4(2)     & 1000  \\
$.250$ &                 &  0.0152(4)    &  14.1(1)     & 1000  \\
\end{tabular}
\caption{ Observables measured on a $12^4$ lattice with four fermi coupling 
$G^2 = 1/4$. Finite size study.}
\end{table}
\narrowtext

\end{document}